\documentclass[12pt,a4paper]{article}
\usepackage{fullpage}
\usepackage[centertags]{amsmath}
\allowdisplaybreaks[4]
\usepackage{amsbsy}
\usepackage{amsfonts}
\usepackage{amssymb}
\usepackage[dvips]{graphicx}
\newcommand{\boss}[2]{\ensuremath{\rlap{\kern-2.5pt\ensuremath{\overset{\scriptscriptstyle(-)}{\phantom{#1}}}}{\ensuremath{{#1}_{#2}}}}}
\usepackage{url}
\usepackage[dvips,hyperindex]{hyperref}
\begin{document}

\begin{flushright}
\begin{tabular}{l}
hep-ph/0610352
\\
3 July 2007
\end{tabular}
\end{flushright}
\vspace{1cm}
\begin{center}
\large\bfseries
Short-Baseline Active-Sterile Neutrino Oscillations?
\\[0.5cm]
\normalsize\normalfont
Carlo Giunti
\\
\small\itshape
INFN, Sezione di Torino,
\\
\small\itshape
and
\\
\small\itshape
Dipartimento di Fisica Teorica,
Universit\`a di Torino,
\\
\small\itshape
Via P. Giuria 1, I--10125 Torino, Italy
\\[0.5cm]
\normalsize\normalfont
Marco Laveder
\\
\small\itshape
Dipartimento di Fisica ``G. Galilei'', Universit\`a di Padova,
\\
\small\itshape
and
\\
\small\itshape
INFN, Sezione di Padova,
\\
\small\itshape
Via F. Marzolo 8, I--35131 Padova, Italy
\end{center}
\begin{abstract}
We suggest the possibility that the
anomalies observed in
the LSND experiment
and
the Gallium radioactive source experiments
may be due to neutrino oscillations generated by a large squared-mass difference
of about $ 20 - 30 \, \text{eV}^2 $.
We consider the simplest 3+1 four-neutrino scheme
that can accommodate also the observed solar and atmospheric neutrino oscillations.
We show that, in this framework,
the disappearance of
$\boss{\nu}{e}$ and $\boss{\nu}{\mu}$
in short-baseline neutrino oscillation experiments
is mainly due to active-sterile transitions.
The implications of the first MiniBooNE results,
appeared after the completion of this paper,
are discussed in an addendum.
\end{abstract}

Neutrino oscillation experiments
have shown that neutrinos are massive and mixed particles
(see the reviews in
Refs.~\cite{Bilenky:1978nj,Bilenky:1987ty,hep-ph/9812360,hep-ph/0202058,hep-ph/0310238,hep-ph/0405172,hep-ph/0506083,hep-ph/0606054}).
The observation of
$\nu_{e}\to\nu_{\mu,\tau}$ oscillations
with a squared-mass difference
\begin{equation}
\Delta{m}^{2}_{\text{SOL}} \simeq 8 \times 10^{-5} \, \text{eV}^{2}
\label{SOL}
\end{equation}
in solar and reactor
neutrino experiments
and the observation of
$\nu_{\mu}\to\nu_{\tau}$ oscillations
with a squared-mass difference
\begin{equation}
\Delta{m}^{2}_{\text{ATM}} \simeq 2.5 \times 10^{-3} \, \text{eV}^{2}
\label{ATM}
\end{equation}
in atmospheric and accelerator
neutrino experiments
can be accommodated in the minimal framework of three-neutrino mixing,
in which the three active flavor neutrinos
$\nu_{e}$,
$\nu_{\mu}$, and
$\nu_{\tau}$
are superpositions of three massive neutrinos
$\nu_{1}$,
$\nu_{2}$, and
$\nu_{3}$.
This three-neutrino mixing framework cannot explain through neutrino oscillations
the LSND
$\bar\nu_{\mu}\to\bar\nu_{e}$
signal
\cite{nucl-ex/9504002,nucl-ex/9605001,nucl-ex/9605003,hep-ex/0104049},
which requires a squared-mass difference
\begin{equation}
\Delta{m}^{2}_{\text{LSND}} \gtrsim 10^{-1} \, \text{eV}^{2}
\,.
\label{LSND}
\end{equation}

Another anomaly observed in neutrino experiments is
the disappearance of $\nu_{e}$'s in the Gallium radioactive source experiments
GALLEX
\cite{Anselmann:1995ar,Hampel:1998fc}
and SAGE
\cite{Abdurashitov:1996dp,hep-ph/9803418,nucl-ex/0512041}.
These experiments are tests of solar neutrino detectors in which
intense artificial ${}^{51}\text{Cr}$ and ${}^{37}\text{Ar}$ neutrino sources
were placed near or inside the detectors.
Both
${}^{51}\text{Cr}$ and ${}^{37}\text{Ar}$ decay through electron capture
($ e^{-} + {}^{51}\text{Cr} \to {}^{51}\text{V} + \nu_{e} $
and
$ e^{-} + {}^{37}\text{Ar} \to {}^{37}\text{Cl} + \nu_{e} $).
The energies of the emitted neutrinos are, respectively,
$ E = 752.73 \pm 0.24 \, \text{keV} $
and
$ E = 813.5 \pm 0.3 \, \text{keV} $
\cite{NuDat}.
The neutrinos emitted by the artificial sources
were detected through the same reaction used for the detection of solar neutrinos
\cite{Kuzmin-Ga-65}:
\begin{equation}
\nu_{e} + {}^{71}\text{Ga} \to {}^{71}\text{Ge} + e^{-}
\,,
\label{j054}
\end{equation}
which has the low neutrino energy threshold
$ E_{\text{th}} = 0.233 \, \text{MeV} $.
The weighted average value of the ratio $R$ of measured and predicted ${}^{71}\text{Ge}$
production rates is
\cite{nucl-ex/0512041}
\begin{equation}
R
=
0.88 \pm 0.05
\,.
\label{001}
\end{equation}
In Ref.~\cite{nucl-ex/0512041} it has been suggested that
this anomaly may be due to an overestimate of the theoretical
cross section of the Gallium detection process in Eq.~(\ref{j054}).
However,
a Gallium cross section rescaled by the factor in Eq.~(\ref{001})
leads to a significant deterioration of the fit of solar neutrino data
\cite{hep-ph/0605186}.

In this paper we consider the possibility that
the anomaly observed in Gallium radioactive source experiments
is due to neutrino oscillations\footnote{
The results of the first GALLEX
artificial ${}^{51}\text{Cr}$ source experiment \cite{Anselmann:1995ar}
has been used in Ref.~\cite{hep-ph/9411414}
in order to constrain the neutrino mixing parameters.
}.

Since the neutrino path in the Gallium radioactive source experiments
was of the order of 10 cm,
an explanation of the observed disappearance of $\nu_{e}$'s through neutrino oscillations
requires a large squared-mass difference
$\Delta{m}^{2}_{\text{Ga}}$.
In fact, requiring an oscillation length $ L_{\text{osc}}^{\text{Ga}} = 4 \pi E / |\Delta{m}^{2}_{\text{Ga}}| $
smaller than about 10 cm,
we obtain
\begin{equation}
\Delta{m}^{2}_{\text{Ga}} \gtrsim 20 \, \text{eV}^{2}
\,.
\label{Ga}
\end{equation}

Assuming CPT invariance,
the survival probability of neutrinos and antineutrinos are equal.
It follows that
the disappearance of electron neutrinos at the level indicated by Gallium radioactive source experiments
appears to be in contradiction with the results of reactor neutrino oscillation experiments
(see the review in Ref.~\cite{hep-ph/0107277}),
which did not observe any disappearance of electron antineutrinos with an average energy of about 4 MeV
at distances between about 10 and 100 m from the reactor source.
Let us notice, however,
that the oscillation length of reactor neutrinos implied by Eq.~(\ref{Ga}) is much shorter than 10 m:
\begin{equation}
L_{\text{osc}}^{\text{reactors}}
\lesssim
40 \, \text{cm}
\,.
\label{010}
\end{equation}
Hence,
in reactor neutrino experiments the oscillations due to
$\Delta{m}^{2}_{\text{Ga}}$
are seen as an energy-independent suppression
of the electron antineutrino flux
by the factor in Eq.~(\ref{001}).
A measurement of such a suppression requires a precise calculation
of the absolute electron antineutrino flux produced in a reactor\footnote{
Information on $\bar\nu_{e}$ disappearance
which is independent of the absolute flux calculation
can be obtained through the measurement of the energy spectrum
(assuming it to be known with small uncertainties)
or the comparison between rates measured with different source-detector distances.
In these cases,
reactor neutrino experiments are not
sensitive to oscillations generated by a squared-mass difference $ \Delta{m}^{2} \gtrsim 2 \, \text{eV}^2 $,
as one can see, for example,
from Fig.~13a of Ref.~\cite{Zacek:1986cu}.
}.
Since this calculation is rather difficult,
it is possible that its systematic uncertainties have been underestimated.
Therefore,
a $\bar\nu_{e}$ disappearance
in reactor neutrino oscillation experiments
at the level indicated by Eq.~(\ref{001})
with the oscillation length in Eq.~(\ref{010})
is not excluded with absolute certainty.

In this paper, we consider the possibility that both the LSND and Gallium anomalies are
due to neutrino oscillations,
through the same large squared-mass difference
\begin{equation}
\Delta{m}^{2}_{\text{LSND}+\text{Ga}} \gtrsim 20 \, \text{eV}^{2}
\,.
\label{LSNDGa}
\end{equation}

We consider, for simplicity,
a four-neutrino mixing scheme,
in which
the three active flavor neutrinos
$\nu_{e}$,
$\nu_{\mu}$,
$\nu_{\tau}$,
and one sterile neutrino
$\nu_{s}$
are superpositions of four massive neutrinos
$\nu_{1}$,
$\nu_{2}$,
$\nu_{3}$, and
$\nu_{4}$.
This is the simplest scheme in which there are three independent squared-mass differences
which can accommodate the hierarchy
\begin{equation}
\Delta{m}^{2}_{\text{SOL}}
\ll
\Delta{m}^{2}_{\text{ATM}}
\ll
\Delta{m}^{2}_{\text{LSND}+\text{Ga}}
\,.
\label{hierarchy}
\end{equation}
Four-neutrino mixing have already been considered in many papers
as the explanation of the LSND anomaly
(see the reviews in Refs.~\cite{hep-ph/9812360,hep-ph/0202058,hep-ph/0405172,hep-ph/0606054}).
Here, we further constrain the allowed values of the large squared-mass difference
and the mixing of the electron neutrino by requiring that
$\Delta{m}^{2}_{\text{LSND}+\text{Ga}}$
is responsible of both the LSND and Gallium anomalies.

Since the so-called 2+2 schemes are disfavored by the data
\cite{hep-ph/0001101,hep-ph/0009299,hep-ph/0011054,hep-ph/0011245,hep-ph/0105269,hep-ph/0112103,hep-ph/0201134,hep-ph/0405172},
we consider a 3+1 scheme,
in which there is a group of three neutrino masses
which is separated from an isolated neutrino mass by the $\text{LSND}+\text{Ga}$
mass splitting.
In this case,
we have
\begin{equation}
\Delta{m}^{2}_{\text{SOL}} = \Delta{m}^{2}_{21}
\,,
\qquad
\Delta{m}^{2}_{\text{ATM}} = |\Delta{m}^{2}_{31}| \simeq |\Delta{m}^{2}_{32}|
\,,
\label{dm2-3}
\end{equation}
\begin{equation}
\Delta{m}^{2}_{\text{LSND}+\text{Ga}} = |\Delta{m}^{2}_{41}| \simeq |\Delta{m}^{2}_{42}| \simeq |\Delta{m}^{2}_{43}|
\,,
\label{dm2-4}
\end{equation}
where
$ \Delta{m}^{2}_{kj} \equiv m_{k}^{2} - m_{j}^{2} $.
Furthermore,
we take into account the upper limit
\begin{equation}
m_{\beta}
<
2.3 \, \text{eV}
\quad
(95\% \, \text{CL})
\,,
\label{n032}
\end{equation}
obtained in
the Mainz \cite{hep-ex/0412056}
and Troitzk \cite{Lobashev:1999tp} tritium $\beta$-decay experiments
on the effective electron neutrino mass \cite{Shrock:1980vy,McKellar:1980cn,Kobzarev:1980nk}
\begin{equation}
m_{\beta}^{2} = \sum_{k=1}^{4} |U_{ek}|^{2} \, m_{k}^{2}
\,.
\label{mbeta}
\end{equation}
Since the three active flavor neutrinos must have large mixings with
$\nu_{1}$,
$\nu_{2}$, and
$\nu_{3}$
in order to accommodate the observed oscillations due to
$\Delta{m}^{2}_{\text{SOL}}$
and
$\Delta{m}^{2}_{\text{ATM}}$,
the only scheme allowed is the one in which
$\nu_{1}$,
$\nu_{2}$, and
$\nu_{3}$
are light, with masses
\begin{equation}
m_{1}
\,,
m_{2}
\,,
m_{3}
\lesssim
2.3 \, \text{eV}
\,,
\label{m123}
\end{equation}
and
$\nu_{4}$ is heavy, with mass
\begin{equation}
m_{4}
\simeq
\sqrt{\Delta{m}^{2}_{\text{LSND}+\text{Ga}}}
\gtrsim
4.5 \, \text{eV}
\,.
\label{m4}
\end{equation}

In four-neutrino schemes,
the average $\nu_{e}$ survival probability in the Gallium experiments is given by
\begin{equation}
\langle P_{\nu_{e}\to\nu_{e}} \rangle
=
1 - \frac{1}{2} \, \sin^2 2\vartheta_{\text{Ga}}
\,,
\label{Pee}
\end{equation}
where $\vartheta_{\text{Ga}}$ is an effective mixing angle.
Interpreting $R$ in Eq.~(\ref{001}) as $\langle P_{\nu_{e}\to\nu_{e}} \rangle$,
we obtain
\begin{equation}
\sin^2 2\vartheta_{\text{Ga}} = 0.24 \pm 0.10
\,.
\label{002}
\end{equation}

In the 3+1 mixing schemes
(see the review in Ref.~\cite{hep-ph/9812360}),
the survival and transition probabilities
in short-baseline neutrino oscillation experiments
have the two-neutrino mixing forms
(for $\alpha,\beta=e,\mu,\tau,s$)
\begin{align}
\null & \null
P_{\nu_{\alpha}\to\nu_{\alpha}}
=
P_{\bar\nu_{\alpha}\to\bar\nu_{\alpha}}
=
1
-
\sin^2 2\vartheta_{\alpha\alpha}
\,
\sin^2 \left( \frac{ \Delta{m}^2_{41} \, L }{ 4 \, E } \right)
\,,
\label{301}
\\
\null & \null
P_{\nu_{\alpha}\to\nu_{\beta}}
=
P_{\nu_{\beta}\to\nu_{\alpha}}
=
P_{\bar\nu_{\alpha}\to\bar\nu_{\beta}}
=
P_{\bar\nu_{\beta}\to\bar\nu_{\alpha}}
=
\sin^2 2\vartheta_{\alpha\beta}
\,
\sin^2 \left( \frac{ \Delta{m}^2_{41} \, L }{ 4 \, E } \right)
\qquad
(\alpha\neq\beta)
\,,
\label{302}
\end{align}
where $L$ is the source--detector distance and
the effective mixing angles are given by
\begin{align}
\null & \null
\sin^2 2\vartheta_{\alpha\alpha}
=
4 \, |U_{\alpha4}|^2 \left( 1 - |U_{\alpha4}|^2 \right)
\,,
\label{311}
\\
\null & \null
\sin^2 2\vartheta_{\alpha\beta}
=
\sin^2 2\vartheta_{\beta\alpha}
=
4 \, |U_{\alpha4}|^2 \, |U_{\beta4}|^2
\qquad
(\alpha\neq\beta)
\,.
\label{312}
\end{align}
Therefore,
we have
\begin{equation}
\sin^2 2\vartheta_{\text{Ga}}
=
\sin^2 2\vartheta_{ee}
=
4 \, |U_{e4}|^2 \left( 1 - |U_{e4}|^2 \right)
\,.
\label{s2t2Ga}
\end{equation}
Taking into account that $|U_{e4}|^2$ is small,
in order to accommodate the observed oscillations due to
$\Delta{m}^{2}_{\text{SOL}}$
and
$\Delta{m}^{2}_{\text{ATM}}$,
we obtain,
from Eqs.~(\ref{002}) and (\ref{s2t2Ga}),
\begin{equation}
|U_{e4}|^2
\simeq
\frac{1}{4} \, \sin^2 2\vartheta_{\text{Ga}}
\simeq
0.06 \pm 0.03
\,.
\label{003}
\end{equation}

In spite of the relatively heavy mass of $\nu_{4}$ in Eq.~(\ref{m4}),
the mixing of $\nu_{e}$ with $\nu_{4}$ is not a problem for
the bound in Eq.~(\ref{n032}) on the effective electron neutrino mass
in $\beta$-decay experiments.
In fact,
the contribution of $\nu_{4}$ to $m_{\beta}$ is
\begin{equation}
m_{\beta}(\nu_{4})
=
|U_{e4}| \, m_{4}
\simeq
1.1 \pm 0.3 \, \text{eV} \left( \frac{ m_{4} }{ 4.5 \, \text{eV} } \right)
\,.
\label{021}
\end{equation}
Therefore, the bound in Eq.~(\ref{n032}) implies
\begin{equation}
m_{4} \lesssim 10 \, \text{eV}
\,.
\label{022}
\end{equation}
Taking into account also Eq.~(\ref{LSNDGa}),
we obtain the allowed range
\begin{equation}
20 \, \text{eV}^{2}
\lesssim
\Delta{m}^{2}_{\text{LSND}+\text{Ga}}
\lesssim
100 \, \text{eV}^{2}
\,.
\label{dm2}
\end{equation}

Let us now consider the LSND
$\bar\nu_{\mu}\to\bar\nu_{e}$
signal, which has been observed with the probability
\cite{hep-ex/0104049}
\begin{equation}
P_{\bar\nu_{\mu}\to\bar\nu_{e}}^{\text{LSND}}
=
\left( 2.64 \pm 0.67 \pm 0.45 \right) \times 10^{-3}
\,.
\label{PLSND}
\end{equation}
Since we are considering large values of $\Delta{m}^{2}_{\text{LSND}+\text{Ga}}$
in the interval in Eq.~(\ref{dm2}),
the transition probability measured in the LSND experiment
is the averaged probability
\begin{equation}
\langle P_{\bar\nu_{\mu}\to\bar\nu_{e}} \rangle
=
\frac{1}{2} \, \sin^2 2\vartheta_{\text{LSND}}
\,,
\label{Pave}
\end{equation}
with the effective mixing angle given by
(see Eq.~(\ref{312}))
\begin{equation}
\sin^2 2\vartheta_{\text{LSND}}
=
\sin^2 2\vartheta_{e\mu}
=
4 \, |U_{e4}|^2 \, |U_{\mu4}|^2
\,.
\label{s2t2em}
\end{equation}
Thus,
from Eq.~(\ref{PLSND}), we obtain
\begin{equation}
\sin^2 2\vartheta_{\text{LSND}}
\simeq
\left( 5.3 \pm 1.6 \right) \times 10^{-3}
\,.
\label{s2t2LSND}
\end{equation}

Short-baseline
$\boss{\nu}{\mu}\to\boss{\nu}{e}$
oscillations generated by
a large squared-mass difference
have been recently searched, with negative results, in the
CCFR \cite{hep-ex/9611013},
KARMEN \cite{hep-ex/0203021},
NuTeV \cite{hep-ex/0203018}, and
NOMAD \cite{hep-ex/0306037} experiments.
From Fig.~8 of Ref.~\cite{hep-ex/0306037},
one can see that
the range of $\sin^2 2\vartheta_{\text{LSND}}$ in Eq.~(\ref{s2t2LSND})
is compatible with the results of the
CCFR,
KARMEN, and
NuTeV experiments if the allowed interval
of $ \Delta{m}^{2}_{\text{LSND}+\text{Ga}} $
in Eq.~(\ref{dm2}) is restricted to
\begin{equation}
20 \, \text{eV}^{2} \lesssim \Delta{m}^{2}_{\text{LSND}+\text{Ga}} \lesssim 30 \, \text{eV}^{2}
\,.
\label{011}
\end{equation}
In fact,
although a combined analysis of all the relevant neutrino oscillations data
yields a poor goodness of fit\footnote{
The fit can be improved by introducing a second sterile neutrino \cite{hep-ph/0305255},
in a so-called 3+2 mixing scheme.
However, it seems to us that it is highly unlikely that
the two large squared-mass differences happen to have
just the right values in the small regions which are not excluded by
the neutrino oscillation data.
}
\cite{hep-ph/9606411,hep-ph/9607372,hep-ph/9903454,hep-ph/0102252,hep-ph/0107150,hep-ph/0112103,hep-ph/0201134,hep-ph/0405172},
if the fit is accepted,
there is an allowed region in the
$\sin^2 2\vartheta_{\text{LSND}}$--$\Delta{m}^{2}_{\text{LSND}+\text{Ga}}$
at
$ \sin^2 2\vartheta_{\text{LSND}} \simeq \left( 2 - 5 \right) \times 10^{-3} $
and
$ \Delta{m}^{2}_{\text{LSND}+\text{Ga}} $
in the interval in Eq.~(\ref{011})
(see
Fig.~3 of Ref.\cite{hep-ph/0505216}
and
Fig.~4 of the first arXiv version of Ref.\cite{hep-ph/0305255}).
This allowed region appears to be in contradiction only with the exclusion curve
obtained in the NOMAD experiment
(see Fig.~8 of Ref.~\cite{hep-ex/0306037}).

From now on, we consider
$\Delta{m}^{2}_{\text{LSND}+\text{Ga}}$
in the interval in Eq.~(\ref{011}).

The range of $\sin^2 2\vartheta_{\text{LSND}}$ in Eq.~(\ref{s2t2LSND})
and
the determination of $|U_{e4}|^2$ in Eq.~(\ref{003}) from the Gallium anomaly
allow us to determine the allowed range of $|U_{\mu4}|^2$:
from Eq.~(\ref{s2t2em}), we obtain
\begin{equation}
|U_{\mu4}|^2
=
\frac{ \sin^2 2\vartheta_{\text{LSND}} }{ 4 \, |U_{e4}|^2 }
\simeq
\frac{ \sin^2 2\vartheta_{\text{LSND}} }{ \sin^2 2\vartheta_{\text{Ga}} }
\simeq
0.02 \pm 0.01
\,.
\label{013}
\end{equation}
This small value of $|U_{\mu4}|^2$ implies that the effective mixing angle in
short-baseline $\nu_{\mu}$ disappearance experiments is given by
\begin{equation}
\sin^2 2\vartheta_{\mu\mu}
\simeq
4 \, |U_{\mu4}|^2
\simeq
0.08 \pm 0.04
\,.
\label{s2t2mm}
\end{equation}
This value of $\sin^2 2\vartheta_{\mu\mu}$
is compatible with the exclusion curves
of the
CDHSW \cite{Dydak:1984zq}
and
CCFR \cite{Stockdale:1985ce}
$\nu_{\mu}\to\nu_{\mu}$
oscillation experiments
for $\Delta{m}^{2}_{\text{LSND}+\text{Ga}}$ in the interval in Eq.~(\ref{011}).
It is interesting to notice that the results of the CDHSW
$\nu_{\mu}$
disappearance experiment favor a $\Delta{m}^{2}_{\text{LSND}+\text{Ga}}$ in the range in Eq.~(\ref{011}),
as remarked at the end of the appendix of Ref.~\cite{hep-ph/0305255}.

Let us now consider the experimental bounds on
$\nu_{\mu}\to\nu_{\tau}$
and
$\nu_{e}\to\nu_{\tau}$
transitions
obtained in short-baseline experiments
(CHORUS \cite{Eskut:2000de},
NOMAD \cite{hep-ex/0306037} and
CCFR \cite{hep-ex/9506007,hep-ex/9809023}).
From Fig.~14 of Ref.~\cite{hep-ex/0106102},
one can see that,
for $\Delta{m}^{2}_{\text{LSND}+\text{Ga}}$
in the range in Eq.~(\ref{011}),
the effective mixing angles
\begin{equation}
\sin^2 2\vartheta_{e\tau} = 4 \, |U_{e4}|^2 \, |U_{\tau4}|^2
\,,
\qquad
\sin^2 2\vartheta_{\mu\tau} = 4 \, |U_{\mu4}|^2 \, |U_{\tau4}|^2
\label{014}
\end{equation}
are bounded by
\begin{equation}
\sin^2 2\vartheta_{e\tau} \lesssim 1 \times 10^{-1}
\,,
\qquad
\sin^2 2\vartheta_{\mu\tau} \lesssim 2 \times 10^{-3}
\,.
\label{0141}
\end{equation}
Taking into account the allowed ranges of
$|U_{e4}|^2$ and $|U_{\mu4}|^2$
in Eqs.~(\ref{003}) and (\ref{013}),
the limit on $\sin^2 2\vartheta_{e\tau}$ does not give a significant bound,
whereas the limit on $\sin^2 2\vartheta_{\mu\tau}$ yields
\begin{equation}
|U_{\tau4}|^2
=
\frac{ \sin^2 2\vartheta_{\mu\tau} }{ 4 \, |U_{\mu4}|^2 }
\simeq
\frac{ \sin^2 2\vartheta_{\mu\tau} \, \sin^2 2\vartheta_{\text{Ga}} }{ 4 \, \sin^2 2\vartheta_{\text{LSND}} }
\lesssim
0.05
\,.
\label{015}
\end{equation}
Therefore, also
$|U_{\tau4}|^2$
is constrained to be small.
It follows that
\begin{equation}
|U_{s4}|^2
=
1 - \left( |U_{e4}|^2 + |U_{\mu4}|^2 + |U_{\tau4}|^2 \right)
\gtrsim
0.8
\,,
\label{016}
\end{equation}
and
the $\nu_{e}$ disappearance indicated by Gallium radioactive source experiments
is mainly due to
$\nu_{e}\to\nu_{s}$
transitions with an effective mixing angle given by
\begin{equation}
\sin^2 2\vartheta_{es}
=
4 \, |U_{e4}|^2 \, |U_{s4}|^2
\simeq
0.2 \pm 0.1
\,.
\label{017}
\end{equation}
These transitions are compatible with the
CCFR bound on $\nu_{e}\to\nu_{s}$
transitions (Fig.~4 of Ref.~\cite{hep-ex/9809023})
for the effective squared mass difference
$\Delta{m}^{2}_{\text{LSND}+\text{Ga}}$
confined in the range in Eq.~(\ref{011}).

The $\nu_{e}\to\nu_{s}$
transitions due to $\Delta{m}^{2}_{\text{LSND}+\text{Ga}}$
affect also solar neutrino experiments.
Since the mixing of $\nu_{s}$ with $\nu_{1}$, $\nu_{2}$, and $\nu_{3}$
is small,
in practice solar neutrino experiments should observe an average probability of
disappearance of electron neutrinos into sterile neutrinos
of the same value as the ratio $R$ in Eq.~(\ref{001})
measured the Gallium radioactive source experiments:
\begin{equation}
\langle P_{\nu_{e}\to\nu_{s}} \rangle
\simeq
\frac{1}{2} \, \sin^2 2\vartheta_{es}
\simeq
0.10 \pm 0.05
\,.
\label{018}
\end{equation}
It is interesting to notice that a comparison of the SNO Neutral-Current (NC) data
with the Standard Solar Model (SSM) prediction is compatible
with $\nu_{e}\to\nu_{s}$ transitions at the level indicated in Eq.~(\ref{018}),
although no evidence can be claimed,
because of the large theoretical uncertainty of the SSM prediction.
In fact,
the equivalent flux of $^{8}\text{B}$ electron neutrinos
measured in SNO\footnote{
The flux in Eq.~(\ref{019})
has been measured in the phase II of the SNO experiment
(also called ``salt phase''),
in which about 2~tons of $\text{Na}\text{Cl}$ have been added to
the heavy water in order to improve the efficiency and precision
of the NC measurement \cite{nucl-ex/0502021}.
}
through the NC reaction
$ \nu + d \to p + n + \nu $,
which is equally sensitive to $\nu_{e}$, $\nu_{\mu}$, and $\nu_{\tau}$,
is \cite{nucl-ex/0502021}
\begin{equation}
\Phi_{\text{NC}}^{\text{SNO}}
=
\left( 4.94 \pm 0.21 {}^{+0.38}_{-0.34} \right) \times 10^{6} \, \text{cm}^{-2} \, \text{s}^{-1}
\,.
\label{019}
\end{equation}
This value can be compared with the
BS05(GS98) \cite{astro-ph/0511337}
and
TC04 \cite{astro-ph/0407176}
SSM values
\begin{align}
\null & \null
\Phi_{{}^{8}\text{B}}^{\text{BS05}}
=
\left( 5.69 {}^{+0.98}_{-0.84} \right) \times 10^{6} \, \text{cm}^{-2} \, \text{s}^{-1}
\,,
\label{121}
\\
\null & \null
\Phi_{{}^{8}\text{B}}^{\text{TC04}}
=
\left( 5.31 \pm 0.6 \right) \times 10^{6} \, \text{cm}^{-2} \, \text{s}^{-1}
\,,
\label{122}
\end{align}
leading to
\begin{align}
\null & \null
\langle P_{\nu_{e}\to\nu_{s}} \rangle_{\text{SNO}+\text{BS05}}
=
1
-
\frac{ \Phi_{\text{NC}}^{\text{SNO}} }{ \Phi_{{}^{8}\text{B}}^{\text{BS05}} }
=
0.13 {}^{+0.15}_{-0.17}
\,,
\label{031}
\\
\null & \null
\langle P_{\nu_{e}\to\nu_{s}} \rangle_{\text{SNO}+\text{TC04}}
=
1
-
\frac{ \Phi_{\text{NC}}^{\text{SNO}} }{ \Phi_{{}^{8}\text{B}}^{\text{TC04}} }
=
0.07 \pm 0.13
\,.
\label{032}
\end{align}
One can see that,
although the uncertainties are large,
the tendency of the ratios in Eqs.~(\ref{031}) and (\ref{032})
is towards an agreement with the average probability of $\nu_{e}\to\nu_{s}$ transitions
in Eq.~(\ref{018}).

The disappearance of $\nu_{e}$ due to $\nu_{e}\to\nu_{s}$ transitions
could affect the search for $\nu_{\mu}\to\nu_{e}$ transitions
in the
MiniBooNE\footnote{
The implications of the first MiniBooNE results,
appeared after the completion of this paper,
are discussed in the addendum at page~\pageref{mb}.
}
experiment \cite{hep-ex/0406048,hep-ex/0407027,hep-ex/0602018},
which has been designed to check the LSND anomaly.
This is due to the fact that the MiniBooNE
$\nu_{\mu}$ beam
has a natural $\nu_{e}$ contamination of about
$ 5 \times 10^{-3}$.
Since the MiniBooNE detector is located at a distance of 541 m from the target and
the energy spectrum of the $\nu_{\mu}$ beam ranges from about 0.2 GeV to about 3 GeV,
with a peak at about 0.6 GeV,
it is convenient to write
the oscillation length due to $\Delta{m}^{2}_{\text{LSND}+\text{Ga}}$ as
\begin{equation}
L_{\text{osc}}
\simeq
120 \, \text{m}
\left( \frac{E}{\text{GeV}} \right)
\left( \frac{\Delta{m}^{2}_{\text{LSND}+\text{Ga}}}{20\,\text{eV}^{2}} \right)^{-1}
\,.
\label{199}
\end{equation}
Hence, a
$\Delta{m}^{2}_{\text{LSND}+\text{Ga}}$
in the range in Eq.~(\ref{011}) implies that the oscillation length is much shorter than the
MiniBooNE source-detector distance and the
flavor transitions
are practically constant over the energy spectrum.
The effect on the $\nu_{e}$ spectrum at the detector
is the superposition of two opposite and competitive contributions:
a $\nu_e$ disappearance due to $\nu_e \to \nu_s$ oscillations
with a relatively large mixing (see Eq.~(\ref{017}))
and
a $\nu_e$ appearance due to  $\nu_\mu \to \nu_e$ with a
relative small mixing (see Eq.~(\ref{s2t2LSND})).
Since the natural contamination of $\nu_e$ in the
$\nu_\mu$ beam is at the percent level, the two opposite
effects on the $\nu_e$ spectrum are competitive.

The hypothesis of
$\nu_{\mu}\to\nu_{e}$ transitions
driven by
$\Delta{m}^{2}_{\text{LSND}+\text{Ga}}$
may soon be tested at the T2K beam line (starting from 2009)
with the near off-axis detector located at a distance of 280 m from the target.
The neutrino energy in T2K is about the same as in MiniBooNE.
With a systematic error on the electron neutrino flux
$\sigma(\text{syst}) \sim 5 \%$ the 90\% C.L. sensitivity to
$\sin^2 2\vartheta_{{\mu}e}$
is about $3 \times 10^{-3}$.

The scenario under consideration
implies also short-baseline $\nu_{\mu}\to\nu_{s}$ oscillations
generated by $\Delta{m}^{2}_{\text{LSND}+\text{Ga}}$
with the effective mixing angle
\begin{equation}
\sin^2 2\vartheta_{{\mu}s}
=
4 \, |U_{\mu4}|^2 \, |U_{s4}|^2
\simeq
0.08 \pm 0.04
\,.
\label{s2t2ms}
\end{equation}
Since this value of
$\sin^2 2\vartheta_{{\mu}s}$
practically coincides with the value of
$\sin^2 2\vartheta_{\mu\mu}$ in Eq.~(\ref{s2t2mm})
and is much larger than the values of
$ \sin^2 2\vartheta_{{\mu}e} = \sin^2 2\vartheta_{\text{LSND}} $
in Eq.~(\ref{s2t2LSND})
and
$ \sin^2 2\vartheta_{\mu\tau} $
in Eq.~(\ref{0141}),
the $\nu_{\mu}\to\nu_{s}$ channel is the dominant cause of
short-baseline $\boss{\nu}{\mu}$ disappearance.

Optimal future experiments which could observe the large disappearance
of $\boss{\nu}{e}$ and $\boss{\nu}{\mu}$ due to active--sterile transitions
and the
$\boss{\nu}{\mu}\leftrightarrows\boss{\nu}{e}$ transitions
due to
$\Delta{m}^{2}_{\text{LSND}+\text{Ga}}$
are:
Beta-Beam experiments \cite{Zucchelli:2002sa}
with a pure $\nu_{e}$ beam from nuclear decay
(see the reviews in Refs.~\cite{physics/0411123,hep-ph/0605033});
Neutrino Factory experiments
with a beam composed of
$\nu_{e}$ and $\bar\nu_{\mu}$,
from $\mu^{+}$ decay,
or
$\bar\nu_{e}$ and $\nu_{\mu}$,
from $\mu^{-}$ decay
(see the review in Ref.~\cite{hep-ph/0210192,physics/0411123});
experiments with a $\bar\nu_{e}$ beam
produced in recoiless nuclear decay
and detected in recoiless nuclear antineutrino capture
\cite{hep-ph/0601079}.

In conclusion,
in this paper we have suggested the possibility that
the anomalies observed in the Gallium radioactive source experiments
and the LSND experiment
may be due to neutrino oscillations generated by the same large squared-mass difference
$\Delta{m}^{2}_{\text{LSND}+\text{Ga}}$.
We have shown that,
in the framework of the simplest 3+1 four-neutrino scheme
that can accommodate also
the $\boss{\nu}{e}\to\boss{\nu}{\mu,\tau}$ oscillations observed in solar and reactor experiments
and
the $\boss{\nu}{\mu}\to\boss{\nu}{\tau}$ oscillations observed in atmospheric and accelerator experiments,
the short-baseline disappearances of $\boss{\nu}{e}$ and $\boss{\nu}{\mu}$
are due mainly to
$\boss{\nu}{e}\to\boss{\nu}{s}$
and
$\boss{\nu}{\mu}\to\boss{\nu}{s}$ transitions,
respectively.
We have noticed that
in the MiniBooNE experiment
flavor transitions are effectively energy-independent
and
the disappearance of $\nu_{e}$ due to $\nu_{e}\to\nu_{s}$ transitions
could affect the search for $\nu_{\mu}\to\nu_{e}$ transitions,
because of the natural $\nu_{e}$ contamination of the beam.
Finally, we have remarked that
the scenario under consideration
could be tested in future experiments with pure $\boss{\nu}{e}$ and $\boss{\nu}{\mu}$ beams,
as Beta-Beam and Neutrino Factory experiments.

\section*{Addendum: First MiniBooNE Results}
\label{mb}

After the completion of this paper,
the MiniBooNE collaboration released their first results
concerning the search for $\nu_{\mu}\to\nu_{e}$
transitions generated by $\Delta{m}^{2}_{\text{LSND}}$
\cite{0704.1500}.
Since no significant excess of quasi-elastic charged-current
$\nu_{e}$ events was observed above the calculated background for
reconstructed neutrino energy $ E_{\nu}^{\text{QE}} > 475 \, \text{MeV} $,
the two-neutrino $\nu_{\mu}\to\nu_{e}$
transitions generated by $\Delta{m}^{2}_{\text{LSND}}$
are disfavored by the MiniBooNE data at 98\% C.L. \cite{0704.1500}.

In the framework of the 3+1 four-neutrino scheme considered in this paper,
the absence of a signal due to
$\nu_{\mu}\to\nu_{e}$ appearance
may be, at least partially,
explained by a suppression of the background
due to $\nu_{e}\to\nu_{s}$ and $\nu_{\mu}\to\nu_{s}$ transitions,
as remarked after Eq.~(\ref{199}).
In fact,
the estimated number of $\nu_{e}$ events is
\begin{equation}
N_{\nu_{e}}
=
P_{\nu_{e}\to\nu_{e}} \, N^{\text{B}}_{\nu_{e}}
+
P_{\nu_{\mu}\to\nu_{\mu}} \, N^{\text{B}}_{\nu_{\mu}}
+
P_{\nu_{\mu}\to\nu_{e}} \, N_{\nu_{\mu}}
\,,
\label{251}
\end{equation}
where
$ N^{\text{B}}_{\nu_{e}} $
and
$N^{\text{B}}_{\nu_{\mu}}$
are, respectively, the estimated numbers of $\nu_{e}$-induced and $\nu_{\mu}$-induced background events,
and
$N_{\nu_{\mu}}$ is the estimated number of $N_{\nu_{e}}$ in the case of full
$\nu_{\mu}\to\nu_{e}$ transmutation.
In short-baseline experiments
$ P_{\nu_{e}\to\nu_{e}} \simeq 1 - P_{\nu_{e}\to\nu_{s}} $,
as remarked after Eq.~(\ref{016}),
and
$ P_{\nu_{\mu}\to\nu_{\mu}} \simeq 1 - P_{\nu_{\mu}\to\nu_{s}} $,
as remarked after Eq.~(\ref{s2t2ms}).
Moreover,
the oscillation probabilities are practically constant
in the MiniBooNE energy spectrum,
as explained after Eq.~(\ref{199}).

From Table~I of Ref.~\cite{0704.1500},
adding the uncertainties in quadrature,
we obtain
\begin{equation}
N^{\text{B}}_{\nu_{e}}
=
229 \pm 32.5
\qquad
\text{and}
\qquad
N^{\text{B}}_{\nu_{\mu}}
=
129 \pm 17.0
\,,
\label{252}
\end{equation}
for $E_{\nu}^{\text{QE}}$
in the range
$ 475 \, \text{MeV} < E_{\nu}^{\text{QE}} < 1250 \, \text{GeV} $.
From the public information kindly given by the MiniBooNE collaboration on the Web\footnote{
\url{http://www-boone.fnal.gov/for_physicists/april07datarelease/}
},
we obtain
\begin{equation}
N_{\nu_{\mu}}
=
62851.2 \pm 250.7
\,.
\label{253}
\end{equation}
From Eqs.~(\ref{001}), (\ref{s2t2mm}) and (\ref{PLSND}), we have
\begin{align}
\null & \null
P_{\nu_{e}\to\nu_{e}} = R = 0.88 \pm 0.05
\,,
\label{254}
\\
\null & \null
P_{\nu_{\mu}\to\nu_{\mu}} = 1 - \frac{1}{2} \, \sin^2 2\vartheta_{\mu\mu} = 0.96 \pm0.02
\,,
\label{255}
\\
\null & \null
P_{\nu_{\mu}\to\nu_{e}} = P_{\bar\nu_{\mu}\to\bar\nu_{e}}^{\text{LSND}} = \left( 2.64 \pm 0.81 \right) \times 10^{-3}
\,.
\label{256}
\end{align}
Then, we obtain
\begin{equation}
P_{\nu_{e}\to\nu_{e}} \, N^{\text{B}}_{\nu_{e}}
=
201.5 \pm 30.8
\,,
\quad
P_{\nu_{\mu}\to\nu_{\mu}} \, N^{\text{B}}_{\nu_{\mu}}
=
123.8 \pm 16.5
\,,
\quad
P_{\nu_{\mu}\to\nu_{e}} \, N_{\nu_{\mu}}
=
165.9 \pm 50.9
\,.
\label{257}
\end{equation}
Comparing with $N^{\text{B}}_{\nu_{e}}$ in Eq.~(\ref{252}),
one can see that the estimated amount of $\nu_{e}$-induced background
is reduced by about 28 events as an effect of $\nu_{e}\to\nu_{s}$ transitions.
This reduction can compensate only partially the larger appearance signal
due to $\nu_{\mu}\to\nu_{e}$ transitions.

The estimated and measured numbers of $\nu_{e}$ events are, respectively,
\begin{equation}
N_{\nu_{e}} = 491.3 \pm 61.7
\qquad
\text{and}
\qquad
N_{\nu_{e}}^{\text{MiniBooNE}} = 380 \pm 19.5
\,,
\label{258}
\end{equation}
Hence,
the 3+1 four-neutrino scheme considered in this paper is compatible with the
results of the MiniBooNE experiment within 1.7 standard deviations.
Although our scheme is clearly not favored by the MiniBooNE data,
we think that further measurements are necessary in order to
assess its viability.

%
%
%
%
%
%
%
%
%
%
%
%


\end{document}